# Stock Price Prediction Using Convolutional Neural Networks on a Multivariate Timeseries

Sidra Mehtab
School of Computing and Analytics
NSHM Knowledge Campus
Kolkata, INDIA
email: smehtab@acm.org

Jaydip Sen
School of Computing and Analytics
NSHM Knowledge Campus
Kolkata, INDIA
email: jaydip.sen@acm.org

*Abstract*—Prediction of future movement of stock prices has been a subject matter of many research work. On one hand, we have proponents of the Efficient Market Hypothesis who claim that stock prices cannot be predicted, on the other hand, there are propositions illustrating that, if appropriately modelled, stock prices can be predicted with a high level of accuracy. There is also a gamut of literature on technical analysis of stock prices where the objective is to identify patterns in stock price movements and profit from it. In this work, we propose a hybrid approach for stock price prediction using machine learning and deep learning-based methods. We select the NIFTY 50 index values of the National Stock Exchange (NSE) of India, over a period of four years: 2015 – 2018. Based on the NIFTY data during 2015 – 2018, we build various predictive models using machine learning approaches, and then use those models to predict the "Close" value of NIFTY 50 for the year 2019, with a forecast horizon of one week, i.e., five days. For predicting the NIFTY index movement patterns, we use a number of classification methods, while for forecasting the actual "Close" values of NIFTY index, various regression models are built. We, then, augment our predictive power of the models by building a deep learning-based regression model using Convolutional Neural Network (CNN) with a walk-forward validation. The CNN model is fine-tuned for its parameters so that the validation loss stabilizes with increasing number of iterations, and the training and validation accuracies converge. We exploit the power of CNN in forecasting the future NIFTY index values using three approaches which differ in number of variables used in forecasting, number of sub-models used in the overall models and, size of the input data for training the models. Extensive results are presented on various metrics for all classification and regression models. The results clearly indicate that CNN-based multivariate forecasting model is the most effective and accurate in predicting the movement of NIFTY index values with a weekly forecast horizon.

*Keywords—Stock Price Prediction, Classification, Regression, Convolutional Neural Network, Multivariate Time Series.*

I. INTRODUCTION

Prediction of future movement of stock prices has been the subject matter of many research work. On one hand, we have proponents of the *Efficient Market Hypothesis* who claim that stock prices cannot be predicted. On the other hand, there are work that have shown that, if correctly modeled, stock prices can be predicted with a fairly reasonable degree of accuracy. The latter have focused on choice of variables, appropriate functional forms and techniques of forecasting. In this regard, Sen and Datta Chaudhauri propose a novel approach of stock price forecasting based on a time series decomposition approach of the stock prices time series [1-8]. Sen proposes a granular approach to stock price prediction in short-term time frame using machine learning and deep learning-based models [9-10].

Mehtab and Sen propose a highly accurate forecasting framework that exploits the power of text mining and natural language prepossessing in analyzing the sentiments in the investment market and utilizing that information in non-linear multivariate predictive framework based on Self-Organizing Fuzzy Neural Networks (SOFNN) [11].

There is also an extent of literature on technical analysis of stock prices where the objective is to identify patterns in stock movements and derive profit from it. The literature is geared towards making money from stock price movements, and various indicators like Bollinger Band, *Moving Average Convergence Divergence* (MACD), *Relative Strength Index* (RSI), *Moving Average, Momentum Stochastics, Meta Sine Wave* etc., have been devised towards this end. There are also patterns like *Head and Shoulders*, *Triangle, Flag, Fibonacci Fan, Andrew's Pitchfork* etc., which are extensively used by traders for gain. These approaches provide the user with visual manifestations of the indicators which helps the ordinary investors to understand which way stock prices may move.

In this paper, we propose several machine learning and deep learning-based predictive models for predicting NIFTY 50 stock price movement in NSE of India. We use daily stock price values for the period January 5, 2015 till December 28, 2018 as the training dataset for the purpose of building the models, and apply the models to predict stock price movement and actual closing value of the stock with a forecast horizon of one week on the test data. For the purpose of testing, we use NIFTY 50 stock price data during the period of December 31, 2018 till December 27, 2019. The prediction framework is further augmented by incorporating powerful deep learning based Convolutional Neural Network (CNN) models for achieving extremely high level of accuracy in prediction of NIFTY 50 index values for the test dataset. Three approaches have been followed in designing CNN models. In the first approach, we build a univariate autoregressive model that predicts the index values for the next week based on the previous "Close" values of the NIFTY index in the training dataset. The second approach follows a multivariate method with each of the variables "Open", "Low", "High", "Close" of the NIFTY index in the training dataset for forecasting the "Close" value of the next week. In the third approach, we build a set of multiple sub-models with each of the four variables being used to construct a separate CNN for univariate forecasting and combining them into a final composite forecast for the "Close" values for the next week.

The rest of the paper is organized as follows. In Section II, we explicitly define the problem at hand. Section III provides a brief review of the related work on stock price movement prediction. In Section IV, we describe our research methodology. Extensive results on the performance of the predictive models are presented in Section V. This section describes the details of all the predictive models built in this work and the results they have produced. Finally Section VI concludes the paper.

## II. PROBLEM STATEMENT

The goal of our work is to collect the stock price of NIFTY 50 from the NSE of India over a reasonably long period of five years, and develop a robust forecasting framework for the forecasting the NIFTY index values. We believe that past movement patterns of daily NIFTY index values can be learned by powerful machine learning and deep learning-based approaches, and that knowledge can be gainfully applied for predicting future movement NIFTY index values. In this work, we choose a prediction horizon of one week and hypothesize that the machine learning-based approaches can be further augmented by powerful deep learning-based methods so that NIFTY index movement can be predicted with even higher accuracy. To validate our hypothesis, in our past work, we used sentiment analysis in the web to build highly accurate predictive models for forecasting future NIFTY index values [11]. In the present work, we follow three different approaches in building Convolutional Neural Networks (CNN) in order to augment to predictive power of our forecasting models. Here, we are not addressing the problem of forecasting of short-term movement of stock price for intra-day trader. Rather, our analysis will be more relevant to a medium-term investor who might be interested in weekly forecast of the NIFTY index values.

At any point of time in the Indian economy, given the appetite of financial market players including individuals, domestic institutions and foreign financial institutions, there is a finite amount of fund that is deployed in the stock market. This amount discounts the entire macroeconomics of that time. This fund would be distributed among various stocks. Thus, on daily basis, if some stock prices are rising, some other stock prices should be falling. Using our proposition, it will be possible for an investor to predict the movement pattern of NIFTY 50 which generally depicts the stock market sentiment in India. The approach builds in indicators like momentum, pivot points and range, all based on daily data on stock prices.

## III. RELATED WORK

The existing propositions in the literature for stock price movement and stock price predictions can be broadly classified into three broad categories based on the choice of variables and approaches and techniques adopted in modeling. The first category includes approaches that use simple regression techniques on cross sectional data [12-16]. These models don't yield very accurate results since stock price movement is a highly non-linear process. The propositions in the second category exploit time series models and techniques using econometric tools like *Autoregressive Integrated Moving Average* (ARIMA), *Granger Causality Test*, *Autoregressive Distributed Lag* (ARDL) and *Quantile Regression* to forecast stock prices [17-20]. The third strand includes propositions using machine learning, deep learning and natural language processing for prediction of stock returns [21-24].

The drawback of the majority of the existing propositions in literature for stock price prediction is their inability to accurately predict highly dynamic and fast changing patterns in stock price movement. The current work attempts to address this shortcoming by exploiting the power of Convolutional Neural Networks in learning the past behavior stock price movements and making a highly accurate forecast for the future behavior of the stock market.

## IV. METHODOLOGY

In Section II, we mentioned that the goal of this work is to develop a framework for daily price movement of NIFTY 50. We collect the NIFTY 50 daily data for the period: January 5, 2015 till December 27, 2019 from the Yahoo Finance website [25]. The raw data consists of the following variables: (i) *Date*, (ii) *Open* value of the index, (iii) *High* value of the index, (iv) *Low* value of the index, (v) *Close* value of the index, and (vi) *Volume* of the stock traded on a given date.

Using the six variables in the raw data, we derive the following variables that we use later on for building our predictive models. We used two approaches in forecasting - regression and classification. The two approaches involved a little difference in using some of the variables, which we will describe and explain later in this Section.

The following nine variables are derived and used in our forecasting models:

a) *month:* it refers to the month to which a given record belongs. This variable is coded into a numeric data, with "1" referring to the month of January and "12" referring to the month of December. The value of the variable *month* lies in the range [1, 12].

b) *day_month:* this variable refers to the day of the month to which a given record belongs. It is a numeric variable lying within the range [1, 31]. For example, the date 14th February 2015 will have a value 14 against the variable *day_month*.

c) *day_week*: it is a numeric variable that refers to the day of the week corresponding to a given record. This variable lies in the range [1, 5]. Monday is coded as 1, while Friday is coded as 5.

d) *close_perc***:** it is a numeric variable that is computed as a standardized value of the percentage change in the *Close* prices on two successive days. The computation of the variable is done as follows. Suppose, we have two successive days: $D_1$ and $D_2$. Let the *Close* price of the stock for $D_1$ is $X_1$ and that for $D_2$ is $X_2$. Then, *close_perc* for $D_2$ computed as $(X_2 - X_1)/X_1$ in terms of percentage.

e) *low_perc:* it is computed in the similar way as *close_norm* as the percentage change in the *Low* values over two successive days.

f) *high_perc*: it is also computed in the same way as the *close_norm* as the percentage change in the *High* values over two consecutive days.

g) *open_perc*: this variable is also computed in the same way as *close_norm* as the percentage change in the *Open* values over successive days.

h) *vol_perc*: computed as the percentage change in *Volume* over two consecutive days.

k) *range_perc*: it is computed as a percentage change in *Range* values for two successive days – the range on a day is computed as the difference between the *High* and the *Low* values on that day.

After, we compute the values of the above nine variables for each day, for the NIFTY 50 data for the period January 5, 2015 – December 27, 2019, we develop the machine learning and deep learning models for classification and regression for predictive modeling.

We use the data for the first four years (i.e., January 5, 2015 – December 28, 2018) for training the models, and test the models using the remaining data, i.e., data for the period December 31, 2018 till December 27, 2019. In regression approach, based on the historical movement of the stock prices, we predict the stock price for a horizon of one week. A week consists of five days - denoted as "mon", "tue", "wed", "thu", and "fri" respectively. In fact, our training data consists of data of 208 weeks and the test data contains data of 52 weeks. We use *close_perc* as the response variable, which is a continuous numeric variable. The objective of the regression technique is to predict the *close_perc* value for each day in the next week, based on the historical data of stock price movement till the current week. If the predicted *close_perc* is positive, then it will indicate that there is an expected rise in the stock price on that day in comparison to its previous day, while a negative *close_perc* will indicate a fall in the stock price on the day.

In the classification approach, the response variable *close_perc* is a categorical variable having two labels. For developing the classification-based modes, we converted *close_perc* into a categorical variable that belongs to one of the two classes – "0" or "1". The value "0" indicating a negative *close_perc* value, while "1" signifying a positive *close_perc* value. Hence, if the forecast model expects a rise in the *close_perc* value on the next day, then the *close_perc* for next day will be "1". A predicted negative value of the *close_norm* on the next day will be indicated by a "0".

For both classification and regression approaches, we experimented with two cases.

**Case I:** We used the data for the period January 5, 2015 till December 28, 2018 that consisted of 1040 daily records for training the model. In other words, training data included data for 208 weeks, where each week consisted of five days. The predictive models are tested on the training data itself to compute accuracy of the models on the training data. Thus, in this case, we make prediction on weekly basis on the training data.

**Case II:** We apply the predictive models on the test data. The test data comprises of data for the period December 28, 2018 till December 27, 2019. Hence, the test data consisted of records for 52 weeks. The predictive power of the models is tested on the test data on prediction horizon of one week.

We use eight approaches of classification and eight approaches of regression for building our forecasting framework. The following classification models are built: (i) *Logistic Regression*, (ii) *K-Nearest Neighbor* (iii) *Decision Tree*, (iv) *Bagging*, (v) *Boosting*, (vi) *Random Forest*, (vii) *Artificial Neural Network*, and (viii) *Support Vector Machines*. The models are tested using the following metrics: (i) sensitivity, (ii) specificity, (iii) positive predictive value, (iv) negative predictive value, and (v) classification accuracy.

In similar line, we designed eight regression models using the following methods: (i) *Multivariate Regression*, (ii) *Decision Tree*, (iii) *Bagging*, (iv) *Boosting*, (v) *Random Forest*, (vi) *Artificial Neural Network*, (vii) *Support Vector Machine*, and (viii) CNN-based multi-step ahead forecasting. Among these approaches, CNN is a deep learning method, while the rest are based on machine learning. For the regression methods, we use Root Mean Square (RMSE), and correlation coefficient between the actual and predicted values of the response variable (i.e., *close_perc*) as the two metrics.

In order to make forecasting robust and more accurate, we train a CNN model. CNNs are proven to be extremely effective in modeling challenging computer vision and image processing problems [26]. In this work, we have explored the power of CNN in forecasting on a complex multivariate time series of NIFTY 50 index values. CNNs have two major types of processing layers – convolutional layers and pooling or subsampling layers. The convolutional layers read an input such as 2-dimensional image or a one-dimensional signal using a kernel by reading the data in small segments at a time and scans across the entire input field. Each read results in an interpretation of the input that is projected onto a filter map and represents an interpretation of the input. The pooling or the subsampling layers take the feature map projections and still them to the most essential elements, such as using a signal averaging (average pool) or signal maximizing process (max pool). The convolution and pooling layers are repeated at depth, providing multiple layers of abstraction of the input signals. The output of the final pooling layer is fed into one or more fully-connected layers that interpret what has been read and maps this internal representation to a class value.

We used the power of CNN in multi-step time series forecasting in the following way. The convolutional layers are used to read sequences of input data and automatically extract features. The pooling layers are used for distilling the extracted features and in focusing attention on the most salient elements. The fully connected layers are deployed to interpret the internal representation and output a vector representing multiple time steps. The benefits that CNN provides in our forecasting job are the automatic feature learning and the ability of the model to output a multi-step vector directly.

We have used CNN in forecasting stock prices in two different ways. In recursive or direct forecast strategy, the model makes one-step predictions and outputs are fed as inputs for subsequent predictions. In the other approach, we used CNNs to predict the entire output sequence as a one-step prediction of the entire vector. Using these two approaches, we build three different types of CNN models for multi-step time series forecasting of stock prices as follows:

(1) Multi-step time series forecasting with univariate input data.

(2) Multi-step time series forecasting with multivariate input data via channels. In this case, each input sequence is read as a separate channel, like different channels of an image (e.g., red, green, blue).

(3) Multi-step time series forecasting with multivariate input data via sub-models. In this case, each input sequence is read by a different CNN sub-model and the internal representations are combined before being interpreted and used to make a prediction.

In the following, we provide brief details of the design of the CNN model for each of the above three cases.

In the first case, we design a CNN for multi-step time series forecasting using only the univariate sequence of the *close_perc* values. In other words, given some number of prior days of "close_perc" values, the model predicts the next standard week of stock market operation. A standard week consists of five days – Monday to Friday. The number of prior days used as input defines the one-dimensional (1D) data of *close_perc* values that the CNN will read and learn for extracting features. There are several choices in deciding on the size and nature of the input to the CNN for training such as: (a) all prior days till the week for which the *close_perc* values to be predicted, (ii) the prior seven days only before the week of prediction, (iii) the prior two weeks (i.e., 10 days, as each week consists of 5 days), (iv) prior one month, (v) the prior week and the week to be predicted in the previous year. Since, there is no obvious best choice here, we have tested the performance of the model on different input sizes and observed the performance of the model under each such cases. Based on the choice of the input, the training data, the test data and prediction process of the model are accordingly designed.

The multi-step time series forecasting approach is essentially an autoregression process. Whether univariate or multivariate, the prior time series data is used for forecasting the values for the next week. With one-week prior data used for building the CNN, we had a small amount of data and hence a very light model. We used only one convolution layer with 16 filters and a kernel size of 3. In other words, it means that the input sequence of five days is read with a convolutional operation in three time steps at a time and this operation is performed 16 times. A pooling layer reduces these feature maps by one-fourth of their size before the internal representation is flattened to one long vector. This is then interpreted by a fully-connected layer before the output layer predicts the *close_perc* values for the next five days. The "ADAM" implementation of the stochastic gradient descent algorithm has been used in the model with 20 epochs and a batch size of 4. For computing the error in prediction, we use RMSE as the metric. With a small batch size and the stochastic nature of the gradient descent algorithm, the model will learn a slightly different mapping of the inputs to the outputs every time it is trained. This implies that performance results may vary in different runs of the model. With the same multi-step univariate autoregressive approach, we increased the number of prior days to 10 days (i.e., two weeks) and trained the model and observed the RMSE of its forecasted values for the next week.

In the second case, i.e., with multi-channel CNN approach, we use each of the four time series variables: *close_perc*, *open_perc*, *high_perc*, and *low_prec* for forecasting the next week's *close_perc* values. We do this by providing each one-dimensional time series to the model as a separate channel of input. In this case, CNN uses a separate kernel and reads each input sequence onto a separate set of filter maps, essentially learning features from each input time series variable. As in the previous case, we modified the way the training and the test data are fed into the model and the way the model is designed. Four input variables are used with two weeks of prior data for the purpose of training the model. The increase in the amount of data requires a larger and more sophisticated model that is trained for longer time. We used two convolutional layers with 32 filter maps followed by a pooling layer, then another convolutional layer with 16 feature maps and pooling. The fully connected layer that interprets the features is increased to 100 nodes and the model is fit for 70 epochs with a batch size of 16 samples.

In the third case, i.e., with multi-step time series forecasting with multivariate input data with sub-models, we have a separate sub-CNN model for each of the four input variable, which we refer to as the multi-headed CNN model. The configuration of the model, including the number of layers and their hyperparameters, are modified to optimize the overall model performance. The multi-headed model is specified using the more flexible functional API for defining Keras models [27]. The program designed for this approach loops over each input variable and creates a sub-model that takes a one-dimensional sequence of 10 days (two weeks) of data and outputs a flat vector containing a summary of the learned features from the sequence. Each of these vectors can are merged by concatenation to make one very long vector that is then interpreted by some fully-connected layers before the forecast of the next week is made. The model needs four arrays as input – one each for the sub-models. We achieved this by creating a list of three-dimensional (3D) arrays, where each 3D array contains [samples, timestamps, 1], with one feature.

## V. Performance Results

In this Section, we provide a detailed discussion on the forecasting models that we have used and the results obtained using those models. We first discuss the classification models, then the machine learning -based regression models and finally the CNN-based deep learning regression model.

For evaluating the classification-based models, we use the following metrics:

*Recall:* It is the ratio of the *true positives (correctly identified "1" s)* to the total number of *positives* in the test dataset expressed as a percentage.

*Specificity:* It is the ratio of the *true negatives (correctly identified "0" s)* to the total number of *negatives* in the test dataset expressed as a percentage.

*Precision:* It is the ratio of the number of *true positives* to the sum of the *true positive* cases and *false positive* cases expressed as a percentage.

*Negative Predictive Value (NPV):* It is the ratio of the number of *true negative* cases to the sum of the *true negative* cases and *false negative* cases expressed as a percentage.

*Classification Accuracy (CA):* It is the ratio of the number of cases which are correctly classified to the total number of cases expressed as a percentage.

For comparing the performance of the regression models, we use the *Root Mean Square Error* (RMSE) values and the product moment correlation values between the predicted and actual *close_perc* values of NIFTY 50. Tables I – VIII depict the performance results of the machine learning-based classification models.

TABLE I. LOGISTIC REGRESSION CLASSIFICATION RESULTS

| Stock | Case I Training Data | | Case II Test Data | |
|---|---|---|---|---|
| NIFTY 50 | Recall | 43.34 | Recall | 73.21 |
| | Specificity | 95.38 | Specificity | 82.02 |
| | Precision | 71.22 | Precision | 61.57 |
| | NPV | 85.21 | NPV | 89.02 |
| | CA | 82.78 | CA | 81.62 |

TABLE II. KNN CLASSIFICATION RESULTS

| Stock | Case I Training Data | | Case II Test Data | |
|---|---|---|---|---|
| NIFTY 50 | Recall | 65.66 | Recall | 21.22 |
| | Specificity | 96.65 | Specificity | 94.98 |
| | Precision | 81.54 | Precision | 45.83 |
| | NPV | 92.12 | NPV | 72.67 |
| | CA | 87.25 | CA | 70.90 |

TABLE III. DECISION TREE (CART) CLASSIFICATION RESULTS

| Stock | Case I Training Data | | Case II Test Data | |
|---|---|---|---|---|
| NIFTY 50 | Recall | 42.88 | Recall | 39.24 |
| | Specificity | 93.21 | Specificity | 89.23 |
| | Precision | 64.82 | Precision | 61.23 |
| | NPV | 84.51 | NPV | 79.21 |
| | CA | 81.23 | CA | 79.49 |

TABLE IV. BAGGING CLASSIFICATION RESULTS

| Stock | Case I Training Data | | Case II Test Data | |
|---|---|---|---|---|
| NIFTY 50 | Recall | 69.21 | Recall | 42.61 |
| | Specificity | 97.36 | Specificity | 87.31 |
| | Precision | 88.51 | Precision | 53.74 |
| | NPV | 91.62 | NPV | 79.42 |
| | CA | 91.24 | CA | 73.72 |

TABLE V. BOOSTING (ADABOOST) CLASSIFICATION RESULTS

| Stock | Case I Training Data | | Case II Test Data | |
|---|---|---|---|---|
| NIFTY 50 | Recall | 100.00 | Recall | 55.21 |
| | Specificity | 100.00 | Specificity | 81.57 |
| | Precision | 100.00 | Precision | 48.62 |
| | NPV | 100.00 | NPV | 85.76 |
| | CA | 100.00 | CA | 74.64 |

TABLE VI. RANDOM FOREST CLASSIFICATION RESULTS

| Stock | Case I Training Data | | Case II Test Data | |
|---|---|---|---|---|
| NIFTY 50 | Recall | 39.46 | Sensitivity | 74.91 |
| | Specificity | 89.15 | Specificity | 64.72 |
| | Precision | 53.42 | Precision | 47.82 |
| | NPV | 83.21 | NPV | 87.51 |
| | CA | 78.42 | CA | 68.34 |

TABLE VII. ANN CLASSIFICATION RESULTS

| Stock | Case I Training Data | | Case II Test Data | |
|---|---|---|---|---|
| NIFTY 50 | Recall | 70.12 | Recall | 21.22 |
| | Specificity | 93.34 | Specificity | 99.72 |
| | Precision | 76.71 | Precision | 99.98 |
| | NPV | 88.23 | NPV | 75.03 |
| | CA | 85.41 | CA | 75.41 |

TABLE VIII. SVM CLASSIFICATION RESULTS

| Stock | Case I Training Data | | Case II Test Data | |
|---|---|---|---|---|
| NIFTY 50 | Recall | 64.71 | Recall | 71.67 |
| | Specificity | 78.53 | Specificity | 75.34 |
| | Precision | 18.13 | Precision | 20.77 |
| | NPV | 96.80 | NPV | 96.72 |
| | CA | 77.58 | CA | 75.03 |

The performance results of the machine learning-based regression models are presented in Tables IX – XV.

TABLE IX. MULTIVARIATE REGRESSION RESULTS

| Stock | Case I Training Data | | Case II Test Data | |
|---|---|---|---|---|
| NIFTY 50 | Correlation | 0.61 | Correlation | 0.57 |
| | RMSE | 33.02 | RMSE | 92.00 |

TABLE X. DECISION TREE REGRESSION RESULTS

| Stock | Case I Training Data | | Case II Test Data | |
|---|---|---|---|---|
| NIFTY 50 | Correlation | 0.98 | Correlation | 0.52 |
| | RMSE | 61.73 | RMSE | 798.41 |

TABLE XI. BAGGING REGRESSION RESULTS

| Stock | Case I Training Data | | Case II Test Data | |
|---|---|---|---|---|
| NIFTY 50 | Correlation | 0.72 | Correlation | 0.55 |
| | RMSE | 24.45 | RMSE | 29.02 |

TABLE XII. BOOSTING REGRESSION RESULTS

| Stock | Case I Training Data | | Case II Test Data | |
|---|---|---|---|---|
| NIFTY 50 | Correlation | 0.71 | Correlation | 0.59 |
| | RMSE | 18.72 | RMSE | 241.36 |

TABLE XIII. RANDOM FOREST REGRESSION RESULTS

| Stock | Case I Training Data | | Case II Test Data | |
|---|---|---|---|---|
| NIFTY 50 | Correlation | 0.95 | Correlation | 0.68 |
| | RMSE | 14.72 | RMSE | 19.26 |

TABLE XIV. ANN REGRESSION RESULTS

| Stock | Case I Training Data | | Case II Test Data | |
|---|---|---|---|---|
| NIFTY 50 | Correlation | 0.71 | Correlation | 0.44 |
| | RMSE | 16.32 | RMSE | 28.72 |

TABLE XV. SVM REGRESSION RESULTS

| Stock | Case I Training Data | | Case II Test Data | |
|---|---|---|---|---|
| NIFTY 50 | Correlation | 0.73 | Correlation | 0.78 |
| | RMSE | 17.31 | RMSE | 15.59 |

We now present the results for the CNN-based regression approach for forecasting the stock prices with a forecast horizon of one week. Table XVI presents the results of multi-step time series forecasting with univariate data containing only the *close_perc* values with a training data size of one week. We found that overall RMSE for the model with a training data input size of 5 (i.e., only one-week prior data), the overall RMSE was 0.895, while RMSE for individual week days were 09., 1.3, 0.7, 0.7, and 0.7 for Monday, Tuesday, Wednesday, Thursday, and Friday respectively. Fig. 1 presents the plot of the RMSE for individual day of the week. The plot exhibited an interesting observation – RMSE for Tuesdays is appreciably higher as compared to the other days in a week. The network design details were discussed in Section IV.

TABLE XVI. CNN REGRESSION RESULTS – UNIVARIATE TIMESERIES WITH PREVIOUS WEEK DATA AS THE TRAINING INPUT

| Stock | Case I Training Data | | Case II Test Data | |
|---|---|---|---|---|
| NIFTY 50 | Correlation | 0.98 | Correlation | 0.96 |
| | RMSE overall | 0.98 | RMSE | 0.90 |
| | Matched Cases | 100% | Matched Cases | 80% |

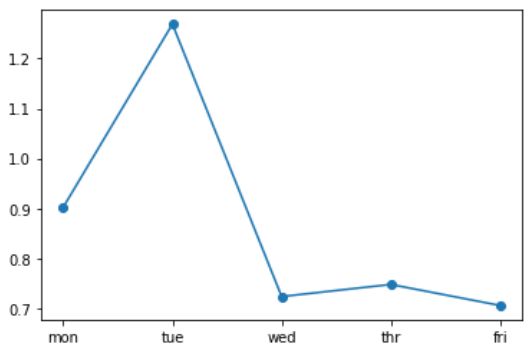

Fig. 1. RMSE with one-week data as input to univariate forecasting

Now, we increased the training data size from 5 to 10, i.e., we provided two weeks' data as an input to the CNN for univariate forecasting. The details of the network design are discussed in Section IV earlier. The overall RMSE for the entire week's forecast was found to be 0.895 with RMSE for the individual days being 0.9, 1.3, 0.7, 0.7, and 0.7 for Monday, Tuesday, Wednesday, Thursday, and Friday respectively. These results were found to be exactly identical to what we observed in with a training data size of 5. However, the matched cases in the test data improved to 100% and correlation coefficient between the forecasted and actual values of open_perc for both the training and test data were found to have increased marginally in this case. The results for this case are presented in Table XVII and Fig. 2.

TABLE XVII. CNN REGRESSION RESULTS – UNIVARIATE TIMESERIES WITH PREVIOUS TWO WEEKS DATA AS THE TRAINING INPUT

| Stock | Case I Training Data | | Case II Test Data | |
|---|---|---|---|---|
| NIFTY 50 | Correlation | 0.99 | Correlation | 0.97 |
| | RMSE overall | 0.98 | RMSE | 0.90 |
| | Matched Cases | 100% | Matched Cases | 100% |

For the second configuration, where we used the CNN model for multi-step forecasting with multiple channel with each channel serving as an input variable, the results obtained were encouraging. We used the network configuration that we discussed earlier in Section IV. The overall RMSE in this case with an input of two weeks' data was found to be 0.4. The individual RMSE were 0.3, 0.4, 0.3, 0.3, 0.2 for Monday, Tuesday, Wednesday, Thursday, and Friday respectively. An appreciable improvement in forecast accuracy was observed in this method using multivariate time series approach as compared to the univariate autoregressive method presented earlier.

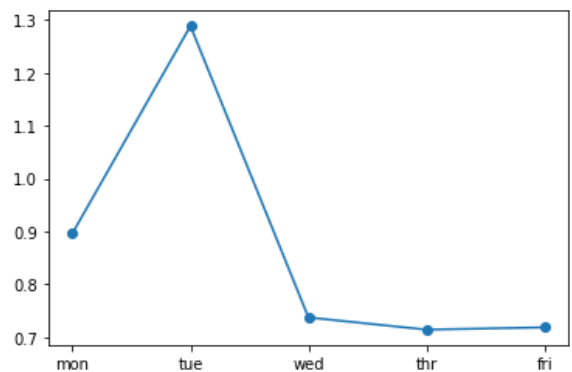

Fig. 2. RMSE with two-weeks' data as input to univariate forecasting

TABLE XVIII. CNN REGRESSION RESULTS – MULTIVARIATE TIMESERIES WITH PREVIOUS TWO WEEKS DATA AS THE TRAINING INPUT

| Stock | Case I Training Data | | Case II Test Data | |
|---|---|---|---|---|
| NIFTY 50 | Correlation | 0.99 | Correlation | 0.97 |
| | RMSE overall | 0.99 | RMSE | 0.93 |
| | Matched Cases | 100% | Matched Cases | 100% |

We now present the results of our third approach to forecasting using CNN models. As discussed in Section IV, this approach involves multi-step time series forecasting with multivariate input data with sub-models. We have a separate sub-CNN model for each of the four input variables- *close_perc*, *open_perc*, *high_perc*, and *low_perc*. As mentioned earlier, in this approach, we have a separate sub-CNN model or head for each input variable, which we refer to as a multi-headed CNN model. The details of design of this network configuration has been presented earlier in Section IV. Table XIX and Figure 3 present the overall performance of this approach. The RMSE for the individual days in a week were found to be 1.2, 1.5, 0.8, 0.9, and 0.9 for Monday, Tuesday, Wednesday, Thursday, and Friday respectively. The overall RMSE for the method was found to be equal to 1.098, which is the highest among all the three approached that we have studied using CNN.

TABLE XIX. CNN REGRESSION RESULTS – MULTIVARITE TIMESERIES WITH EACH VARIABLE BEING USED FOR BUILDING A SEPARATE CNN MODEL WITH THE WITH TWO WEEKS DATA AS THE TRAINING INPUT

| Stock | Case I Training Data | | Case II Test Data | |
|---|---|---|---|---|
| NIFTY 50 | Correlation | 0.99 | Correlation | 0.97 |
| | RMSE overall | 0.98 | RMSE | 1.09 |
| | Matched Cases | 100% | Matched Cases | 100% |

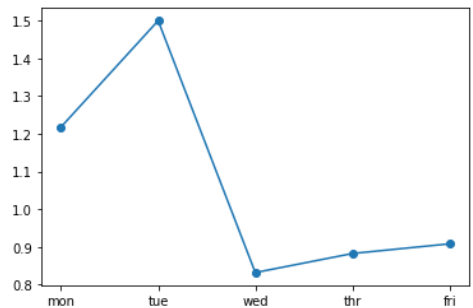

Fig. 3. RMSE with two-weeks' input data to multivariate model with each variable being used to model a separate CNN

Among the classification methods that we have used, the training accuracy for boosting, quite expectedly, was found to have yielded the best results. However, for the test dataset, Random Forest produced the overall best results. Since the NIFTY 50 data was imbalanced with majority number of records belonging to Class "0", ANN performed extremely well on specificity, and precision, while it worked very poorly on recall. boosting, quite expectedly, performed the best for the training dataset. SVM performed quite well, on the average, on all metrics except for precision.

Among the regression models, all the three CNN-models outperformed the machine learning-based approaches. However, among the three CNN models, the one that exploited multi-variate features of the training dataset with walkover cross validation (i.e., our second approach in CNN) yielded the maximum forecast accuracy on the test dataset.

## VI. CONCLUSION

In this paper, we have presented several approaches to stock price and movement prediction on a weekly forecast horizon using eight regression and eight classification methods. These models are based on machine learning and deep learning approaches. We built, fine-tuned, and then tested these models using daily historical data of NIFTY 50 during January 5, 2015 till December 27, 2019. The raw data is suitably pre-processed and suitable variables are identified for building predictive models. After designing and testing the machine learning and deep learning-based models, the predictive framework is further augmented building three Convolutional Neural Network models with univariate and multivariate approaches with varying input data size and network configurations. The performance of these CNN-based deep learning models was found to have far too superior to that of the machine-learning based predictive models. The study has conclusively elicited that fact that deep learning-based models have much higher capability in extracting and learning the features of a training dataset than their corresponding machine learning counterparts. It also reveals the fact that multivariate analysis enables building more accurate forecasting models than univariate analysis. We would explore the possibility of exploiting the power of Generative Adversarial Networks (GAN) in forecasting stock price movement as a future scope of our work.